\font\xmplbx = cmbx10 scaled \magstep2
\font\eightrm=cmr8

\def\pmb#1{\setbox0=\hbox{#1}\kern-.025em
    \copy0\kern-\wd0\kern.05em\kern-.025em\raise.029em\box0}
\def\bsigma{\pmb{$\sigma$}}

\def\a{\alpha}      \def\l{\lambda}   
       \def\m{\mu}
\def\g{\gamma}      \def\G{\Gamma}    \def\n{\nu}
\def\d{\delta}      \def\D{\Delta}         
\def\e{\epsilon}                      
                  \def\s{\sigma}  
                         
        \def\vphi{\varphi}
                     
\def\k{\kappa}

\def\section#1{
  \vskip.7cm\goodbreak
  \noindent{\bf #1}
  \nobreak\vskip.4cm\nobreak  }
  
\def\sqr#1#2{{\vcenter{\hrule height.#2pt
             \hbox{\vrule width.#2pt height#1pt \kern#1pt
             \vrule width.#2pt}
             \hrule height.#2pt}}}
\def\boxx{\hskip1pt\mathchoice\sqr54\sqr54\sqr{3.2}4\sqr{2.6}4
                                                       \hskip1pt}
\def\RR{{\buildrel {\scriptstyle \hphantom{i}o}\over R}\vphantom{R}}
\def\GG{{\buildrel {\scriptstyle \hphantom{i}o}\over G}\vphantom{G}}

\magnification=\magstep1
\baselineskip = 1.5\baselineskip
\parindent = 3em
\hoffset = .2in
\hsize = 5.1in
\null

\vglue 1cm

\centerline {\xmplbx Higher-dimensional geometric $\bsigma$-models}
\vskip 1cm
\centerline{M Vasili\'c \footnote\dag{\eightrm E-mail:
mvasilic@rt270.vin.bg.ac.yu}}
\vskip .3cm
\centerline{\it Department of Theoretical Physics, Institute "Vin\v ca",
             P.O.Box 522,}
\centerline{\it 11001 Beograd, Yugoslavia}
\vskip 2.5cm
\noindent{\bf Abstract.}\hskip .4cm
Geometric $\s$-models have been defined as purely geometric theories of
scalar fields coupled to gravity. By construction, these theories possess
arbitrarily chosen vacuum solutions. Using this fact, one can build a
Kaluza--Klein geometric $\s$-model by specifying the vacuum metric of the
form $M^4\times B^d$. The obtained higher dimensional theory has vanishing
cosmological constant but fails to give massless gauge fields after the
dimensional reduction. In this paper, a modified geometric $\s$-model is
suggested, which solves the above problem.
\vskip 1cm
\noindent PACS numbers: 0240, 0420, 0490

\vfill\eject

\section{1. Introduction}

The cosmological constant problem of Kaluza--Klein theories [1]
whose internal manifold is not Ricci-flat is a longstanding one. In the
conventional Kaluza--Klein treatment, the internal manifold is chosen
in such a way that its isometries define internal symmetries of the theory.
At the same time, a ground state in the form of the direct product of
the 4-dimensional flat spacetime with a compact, nonflat internal space
does not satisfy classical Einstein--Hilbert equations of motion. The attempt
to solve this problem by adding a cosmological term has failed. Indeed,
to reproduce some known gauge couplings, the cosmological constant
is constrained to be of the order of the Planck mass squared, which
strongly disagrees with the observed universe.

Among a variety of existing approaches to this problem, we shall focus
our attention on those which use matter fields to trigger spontaneous
compactification. At the same time, we do not want to lose the geometric
character of our theory. How can we reconcile these two requirements?
Notice, in this respect, that so called geometric $\s$-models have been
defined [2] as purely geometric theories of scalar fields coupled to
gravity. By construction, these scalar fields originate from the
coordinates of the spacetime, and, as a consequence, can be gauged away.
In the context of higher dimensional theories, such an approach has
already been used in literature. The authors of references [3] and [4]
have employed scalar fields in the form of a nonlinear $\s$-model to
trigger the compactification. It is not difficult to see that their
model is a particular example of a geometric $\s$-model. It turns out,
however, that, although solved the cosmological constant problem,
it failed to give massless gauge fields. We shall try to modify the
model of [3] and [4] in the spirit of [2], and reconcile the
masslessness of the gauge fields with the zero value of cosmological
constant. In the course of our analysis, it will become obvious that
more than one model of the kind can be defined.

The lay-out of the paper is as follows. In section 1, we shall analyse
the model suggested in [3] and [4] from the point of view of
geometric $\s$-models. We shall readily use the gauge freedom, and fix
all the scalar fields in the theory thereby reducing it to the purely
geometric, non-covariant equations of motion of the form $R_{MN}=\RR_
{MN}$. The function $\RR_{MN}$ on the right-hand side is the Ricci
tensor of the Kaluza--Klein vacuum $M^4\times B^d$. The linearized
version of the theory turns out to be easily averaged over the
internal coordinates. The resulting effective 4-dimensional equations
of motion are then shown to necessarily contain massive terms in the
sector of gauge fields. A careful analysis of the problem suggests
a simple modification of this model. In section 3, we shall see that
the Yang--Mills sector of the effective 4-dimensional equations of
motion obtained by averaging the equations $R^{MN}=\RR^{MN}$ contains
no massive fields. The fact that the new model uses Ricci tensors
with upper indices turns out to be crucial. However, it makes it
difficult to construct the corresponding Lagrangian. This is why we
suggest another modification of the model in section 4. By adding
terms proportional to $(G_{MN}-\GG_{MN})$ to our equations of
motion, we shall certainly not lose the good property of our
model to have vanishing cosmological constant. Indeed, such
equations possess the vacuum solution $G_{MN}=\GG_{MN}$, and we
choose the metric $\GG_{MN}$ to be of the Kaluza--Klein type
$M^4\times B^d$. We shall be able to demonstrate that, after the
dimensional reduction, our higher dimensional Lagrangian gives
the standard Einstein, Yang--Mills and Klein--Gordon sectors. The
scalar excitations of the internal manifold turn out to be all
massive, with masses of the order of the Planck mass. The
analysis is confined within the linear approximation of the theory.

In section 4, we shall covariantize our model. By employing a set
of $4+d$ scalar fields, a generally covariant $\s$-model of a
non-standard type is obtained. The Lagrangian turns out to be a
non-polynomial function of the scalar field derivatives. We shall
still be able to bring it to a polynomial
form by introducing a set of auxiliary fields. The obtained theory
retains a purely geometric character since all the fields except
$G_{MN}$ are either auxiliary or gauge degrees of freedom.
A brief comparison of the new model with conventional nonlinear
$\s$-models points out some conspicuous differences.

Section 5 is devoted to concluding remarks.

\section{2. Compactification induced by scalars}

The model the authors of references [3] and [4] discuss consists
of Einstein gravity in $4+d$ dimensions coupled to a nonlinear
$\s$-model:
$$
I = - \k^2\int d^{4+d}X\sqrt{-G}\left[ R - F_{ij}(\Omega)\,
\Omega^i_{\,,M}\,\Omega^j_{\,,N}\,G^{MN} \right]\ .
                                                      \eqno(1)
$$
The scalar fields $\Omega^i$, $i=1,2,...,d\,$, are thought of as
coordinates of a d-dimensional compact Riemannian manifold $B^d$
with Ricci tensor $F_{ij}(\Omega)$, while the coordinates $X^M
\equiv (x^{\mu}, y^m)$ parametrize a $(4+d)$-dimensional spacetime
with metric $G_{MN}$. The indices run as follows:
$$
\eqalign{
 M,N      &= 0,1,...,3+d  \cr
 \mu ,\nu &= 0,1,...,3    \cr
 m,n      &= 1,2,...,d \ \ .    }
$$
Notice that the number of scalar fields equals the number of
compact dimensions of the spacetime. The equations of motion
for this theory possess a vacuum solution of the form
$$
G_{MN} = \GG_{MN} \, \qquad \Omega^m = y^m           \eqno(2)
$$
where
$$
\GG_{MN}\equiv \pmatrix{\eta_{\mu\nu}&0\cr
0&\phi_{mn}(y)\cr}\                                 \eqno (3)
$$
and $\phi_{mn}(y)$ stands for the metric of $B^d$. The scalar
sector of the solution (2) is obviously topologically nontrivial
since it is described by a degree one mapping from $B^d$ to
$B^d$. At the same time, the metric (3) has the form of the
direct product of the 4-dimensional Minkowski spacetime with
a compact internal space, as desired. If we restrict our attention
to the physics of small excitations of this vacuum, we can always
choose the spacetime coordinates to fix $\Omega^m = y^m$. Then, the
action functional (1) reduces to
$$
I = - \k^2\int d^{4+d}X\sqrt{-G}\left( R_{MN}-\RR_{MN} \right)
      G^{MN} \ .
                                                      \eqno(4)
$$
where $\ \RR_{MN}\ $ stands for the Ricci tensor of the vacuum
metric (3). This is a purely
\eject\noindent
geometric but non-covariant theory
whose physical content is fully contained in the equatons of
motion
$$
R_{MN} = \RR_{MN} \ .                                     \eqno(5)
$$
Comparing it with the results of [2], we can see that the above
theory is a particular example of a geometric $\s$-model based
on the vacuum metric (3). The covariantization of the equations
(5) in the spirit of [2] introduces $4+d$ scalar fields in the
form of a nonlinear $\s$-model. It is owing to the special form
of $\GG_{MN}$ as given by (3) that only $d$ out of $4+d$ scalar
fields survive.

As mentioned in the introduction, the model (1) solves the
cosmological constant problem but fails to give massless
Yang--Mills fields after the dimensional reduction. When
rewritten as (5), the theory authomatically takes care of the
cosmological term. Indeed, the metric $\GG_{MN}$ is by
definition a solution to the equations of motion (5). To
analyse the spectar of the corresponding effective 4-dimensional
theory, we shall use the standard $4+d$ decomposition [5] of the
metric $G_{MN}$:
$$
G_{MN}\equiv\pmatrix{g_{\mu\nu}+B^k_{\mu}B^l_{\nu}u_{kl}
&B^k_{\mu}u_{kn}\cr B^k_{\nu}u_{km}&u_{mn}\cr}\ .       \eqno (6)
$$
For the perturbations of the vacuum $\GG_{MN}$ we adopt the
notation
$$
g_{\mu\nu}=\eta_{\mu\nu}+h_{\mu\nu} \ \ \ \qquad
                  u_{mn}=\phi_{mn}+\varphi_{mn}\ .       \eqno(7)
$$
Substituting these expressions into (5) we find
$$
\eqalign{
&R^{\mu}_{\ \nu} = 0   \cr
&R^{\mu}_{\ n} = - \RR_{nl}\, B^{l\mu}+O(2)  \cr
&R^m_{\ n} = \RR^m_{\ n}-\RR_{nl}\, \varphi^{ml}+O(2)   }  \eqno(8)
$$
where $\RR_{mn}$ is the d-dimensional Ricci tensor of the vacuum
metric $\phi_{mn}$. For our purposes, the mixed components
$R^M_{\ N}$ of the Ricci tensor turn out to be more convenient.
Indeed, it is the mixed components which, after the decomposition
(6) is employed, give the standard Einstein and Yang--Mills terms.
The expressions on the right-hand
\eject
\noindent
side of (8) then measure the
deviation of our theory from the standard case.
In particular, we shall see that the term $\RR_{nl} B^{l\mu}$ is
responsible for the appearance of massive gauge fields.

To obtain effective 4-dimensional equations of motion, we shall
average the equations (8) over the internal coordinates. The
average of a d-scalar S, as defined by
$$
\langle S \rangle \equiv {{\int d^dy\sqrt{-u}\ S}\over
{\int d^dy\sqrt{-u}}} =
{{\int d^dy\sqrt{-\phi}\ S}\over{\int d^dy\sqrt{-\phi}}}+O(2)
                                                          \eqno(9)
$$
is also a d-scalar. However, a simple definition of the kind for
d-vectors and d-tensors does not exist. This is why we have to
project the equations (8) on an appropriately chosen basis in
$B^d$ before we use (9) to average them. (This kind of dimensional
reduction has already been used in [6] to obtain effective
4-dimensional equations of motion out of dimensionally continued
Euler forms.) As is customary, let us suppose that $B^d$ is a
homogeneous space with $m$ Killing vectors $K^l_a(y)$,
$a=1,...,m\ $, which form a (generally overcomplete) basis in $B^d$.
Using the decomposition
$$
B^m_{\nu}=K^m_aA^a_{\nu}\ \qquad \varphi_{mn}=K_{am}
K_{bn}  \varphi^{ab}
$$
and projecting the vector and tensor equations (8) on the Killing
basis, we obtain an equivalent set of d-scalar field equations.
Although basically linear, the averaged equations will contain
terms of the type $\langle S_1S_2 \rangle$ owing to the presence
of $y$-dependent coefficients. The average $\langle S_1S_2 \rangle$
cannot generally be expressed in terms of $\langle S_1 \rangle$
and $\langle S_2 \rangle$. However, our internal manifold is of the
Planckian size, and it is not unreasonable to restrict our analysis
to solutions which slowly vary in $y$ directon. In that case, the
product of averages $\langle S_1 \rangle \langle S_2 \rangle$
becomes the leading term in the decomposition
$$
\langle S_1S_2 \rangle = \langle S_1 \rangle
\langle S_2 \rangle + \Delta_{12}
$$
so that $\Delta_{12}$ can be regarded as a small correction.
Using this fact and the fact \break
that averages of covariant d-divergences \ vanish, we obtain
the \ following \ effective
\eject\noindent
4-equations:
$$\bar {\cal R}_{\mu\nu}+{1\over 2}\,\bar {\varphi}_{,\mu\nu} =
\emptyset $$
$${\g}_{ab}\,\partial_{\n}\bar F^{b\mu\nu} +
  2\,m_{ab}\,\bar A^{b\mu} = \emptyset                     \eqno(10)  $$
$${\s}_{abcd}\,\boxx \,\bar{\varphi}^{cd} +
  {\mu}_{abcd}\,\bar{\varphi}^{cd} = \emptyset \ . $$
Here, ${\cal R}_{\mu\nu}$ is the Ricci tensor of the 4-metric
$g_{\mu\nu}$, $\boxx$ is the corresponding d'Alember-\break tian, and
$F^a_{\mu\nu}\equiv A^a_{\mu ,\nu}-A^a_{\nu ,\mu}+O(2)$ is
the gauge field strength for the gauge fields $A^a_{\mu}$.
The bar over a quantity denotes its expectation value as defined
by (9), and $\emptyset \equiv \D + O(2)$. The coefficients in (10) 
are vacuum expectation values of products of the Killing vectors
and their covariant derivatives. For example,
$$
\eqalign{
{\g}_{ab}&\equiv \langle K^m_a K_{bm}\rangle \qquad\quad
m_{ab}\equiv \langle K^m_a K^n_b {\RR}_{mn}\rangle    \cr
&{\s}_{abcd}\equiv \langle K^m_a K^n_b K_{cm}K_{dn}
\rangle\  +\  a \leftrightarrow b \ . }
$$
We see that the fields $\bar A^a_{\mu}$ are generally massive
with masses proportional to the curvature of the internal
manifold $B^d$. So are the scalar excitations $\bar {\vphi}^{ab}$,
with the exception of $\bar{\vphi}\equiv {\g}_{ab}\bar{\vphi}^{ab}
= \langle \vphi^m_m \rangle + \D$ which turns out to be massless.
We can use that fact to rescale the metric $\bar g_{\mu\nu}$
according to $\tilde g_{\mu\nu}\equiv (1+\bar{\vphi}/2)\,\bar
g_{\mu\nu}$, whereby the first equation (10)  takes the standard
Einstein form $\tilde{\cal R}_{\mu\nu}=\emptyset$.

\section{3. Massless gauge fields}

By analysing the equations (8), one finds that the term
$\RR_{nl}B^{l\mu}$ on the right-hand side makes the gauge
fields massive. The simplest way to get rid of it is to
postulate the equations of motion of the form $R^M_{\ N}
= \RR^M_{\ N}$. These, however, are not symmetric, and can
only be used within the vielbein formalism. For this reason,
we shall concentrate our attention on the symmetric field
equations of the form
$$
R^{MN} = \RR^{MN} \ .                              \eqno(11) 
$$
It is not difficult to show that the corresponding theory
contains no massive gauge fields. Indeed, by rewriting the
equations (11)  in terms of the mixed components of the
Ricci tensor, we find
$$
\eqalign{
&R^{\mu}_{\ \nu} = 0   \cr
&R^{\mu}_{\ n} = 0    \cr
&R^m_{\ n} = \RR^m_{\ n}+\RR^{ml}\,\varphi_{nl}\ .}  \eqno(12) 
$$
The critical term on the right-hand side of the second equation
(12)  is missing! The effective 4-dimensional equations of
motion are obtained by the exact procedure described in
section 2. The result is
$$\bar {\cal R}_{\mu\nu}+{1\over 2}\,\bar {\varphi}_{,\mu\nu} =
\emptyset $$
$$\partial_{\nu}\bar F^{\ \mu\nu}_a = \emptyset     \eqno(13)  $$
$${\s}_{abcd}\,\boxx \,\bar{\varphi}^{cd} +
  \mu^{\prime}_{abcd}\,\bar{\varphi}^{cd} = \emptyset  $$
where the group metric $\g_{ab}$ is used to raise and lower
the group indices. Comparing it to (10) , we see that there
are no massive terms in the Yang--Mills sector. In addition,
the formerly massless scalar field $\bar {\vphi}$ acquires
mass of the order of the Planck mass. In particular, if we
choose our $B^d$ to be an Einstein manifold, say $\RR_{mn}
=\l \phi_{mn}$, we shall find $(\boxx - 4\l )\bar{\varphi}
=\emptyset$. As a consequence of $\boxx \bar{\varphi}\ne 0$,
the local rescalings of the metric $\bar g_{\mu\nu}$ cannot
bring the first equation (13)  into the standard Einstein form.
Still, it is possible to fix the gauge $\partial^{\nu}\psi_
{\mu\nu}={1\over 2}\bar \varphi_{,\mu}$ in the linearized
theory ($\psi_{\mu\nu}\equiv \bar h_{\mu\nu}-{1\over 2}
\eta_{\mu\nu}\bar h^{\l}_{\l}$) thereby reducing
$\bar {\cal R}_{\mu\nu}+{1\over 2}\bar {\varphi}_{,\mu\nu}
=0$ to $\boxx \bar h_{\mu\nu}=0$, as is customary. In this
respect, notice that, although the equations (11)  are
basically non-covariant, they still possess a partial gauge
symmetry as a consequence of our special choice of $B^d$.
Indeed, it is not difficult to check that the coordinate
transformations
$$
x^{\mu^{\prime}}=x^{\mu^{\prime}}(x) \qquad
y^{m^{\prime}}=y^m + \e^a(x)\,K^m_a(y)
$$
do not change the form of the equations of motion (11) .

Before we covariantize the non-covariant field equations (11) ,
we would like to define the corresponding action functional.
It turns out, however, that no obvious generalization of (4)
exists. In the next section, we shall suggest a Lagrangian
whose equations of motion differ from (11)  but retain all
their good features.

\section{4. Lagrangian}

The geometric $\s$-model approach to the cosmological constant
problem does not uniquely single out the equations of motion
in the form of (5) or (11) . One can always add terms
proportional to $(G_{MN}-\GG_{MN})$ without losing the vacuum
solution $G_{MN}=\GG_{MN}$. We shall use this freedom to
modify the equations (11)  in a way which will allow for a
simple construction of the corresponding Lagrangian. At the
same time, we have to carefully choose this correction in order
not to lose the needed masslessness of the Yang--Mills sector.
A simple analysis along these lines takes us to the following
non-covariant action functional
$$
I = - \k^2\int d^{4+d}X\sqrt{-G}\left[ R-\RR+\RR^{MN}
      \left( G_{MN}-\GG_{MN} \right)\right]\ .
                                                      \eqno(14) 
$$
Varying it with respect to $G_{MN}$ gives the equations of
motion of the form
$$
R^{MN}=\RR^{MN}-{2\over{2+d}}\,G^{MN}\RR^{LR}
\left( G_{LR}-\GG_{LR} \right)\ .                     \eqno(15) 
$$
As we can see, the correction to (11)  is indeed proportional to
$( G_{MN}-\GG_{MN})$. The Yang--Mills sector of the theory is
best analysed if we rewrite (15)  using mixed components of the
Ricci tensor. Then, the equations of motion read
$$
\eqalign{
&R^{\mu}_{\ \nu}=-{2\over{2+d}}\ \d^{\mu}_{\ \nu}\,
                   \RR^{ij}\,\vphi_{ij}   \cr
&R^{\mu}_{\ n} = 0    \cr
&R^m_{\ n} = \RR^m_{\ n}+\RR^{ml}\,\varphi_{nl}
-{2\over{2+d}}\ \d^m_{\ n}\,\RR^{ij}\,
                                \vphi_{ij} \ .}       \eqno(16) 
$$
As in (12) , the crucial term on the right-hand side of the
second equation (16)  is missing. The effective 4-dimensional
equations of motion are obtained using the averaging procedure
of section 2. To simplify the analysis, we shall choose our
internal space in the form of an Einstein manifold
$$
\RR_{mn}=\l\,\phi_{mn}
$$
with $\l<0$ \ in accordance with the adopted conventions
($\ R^M_{\ NLR}=\G^M_{\ NL,R}-\cdots \ $,
\eject\noindent
$diag(G_{MN})=(-,+,...,+)\ $). Then, the averaged equations become
$$
\bar {\cal R}_{\mu\nu}+{1\over 2}\,\bar {\varphi}_{,\mu\nu}
+{{2\l}\over{d+2}}\,\eta_{\m\n}\,\bar\vphi=\emptyset
$$
$$
\partial_{\nu}\bar F^{\ \mu\nu}_a = \emptyset         \eqno(17) 
$$
$$
{\s}_{abcd}\,\boxx \,\bar{\varphi}^{cd} +
\mu^{\prime\prime}_{abcd}\,\bar{\varphi}^{cd}=\emptyset\ .
$$
The gauge fields $\bar A^a_{\m}$ are obviously massless, but
the scalar excitations $\bar \vphi^{ab}$ have masses of the
order of the Planck mass. In particular, the scalar field
$\bar\vphi$, appearing in the first equation (17) , satisfies
$$
\left(\boxx - {{8\l}\over{d+2}}\right)
\bar\varphi=\emptyset\ .                           \eqno(18) 
$$
We see that the conventional choice $\l<0$
ensures the correct sign for the mass term in (18) . Moreover,
as opposed to the case of section 3, the equation (18)  makes
it possible to rescale the metric $\bar g_{\m\n}$ according
to
$$
\tilde g_{\m\n}\equiv \left( 1+{{\bar\vphi}\over 2}\right)
\,\bar g_{\m\n}+O(2)
$$
thereby bringing the first equation (17)  into the standard
Einstein form
$$
\tilde{\cal R}_{\m\n}=\emptyset\ .
$$
The masses $\m^{\prime\prime}_{abcd}\,$, as well as the
coefficients $\s_{abcd}$ and $\g_{ab}\,$, are defined as
vacuum expectation values of products of the Killing vectors
and their covariant derivatives. They are constant tensors
of the isometry group of the internal manifold $B^d$. In the
case of $B^d=S^2$, for example, one finds
$$
\g_{ab}={2\over 3}\,\d_{ab}
$$
$$
\s_{abcd}={2\over {15}}\,\d_{ab}\d_{cd}+{7\over{15}}
\,(\d_{ac}\d_{bd}+\d_{bc}\d_{ad})\ .
$$
The $SO(3)$ tensor $\s_{abcd}$ has the inverse defined through
$(\s^{-1})^{abcd}\s_{cdef}\equiv \d^a_{(e}\,\d^b_{f)}$. As a
consequence, all the scalar fields $\bar\vphi^{ab}$ survive
as independent degrees of freedom in this theory. This is an
improvement as compared to [6] where the cosmological constant
problem has been solved at the expense of losing the kinetic
terms of some
\eject\noindent
scalar excitations. The mass matrix
$\m^{\prime\prime}_{abcd}\,$, being a constant $SO(3)$ tensor
itself, has the same structure as $\s_{abcd}\,$, but requires
a lengthier calculation.

\section{5. Covariantization}

To covariantize the theory given by the action functional (14) ,
we shall follow the ideas of reference [2]. Like there, we
shall use a new set of coordinates, $\Omega^A=\Omega^A(X)$,
$A=0,1,...,3+d\,$, to fix the vacuum quantities of our model.
Then, the covariantization is achieved through the substitution
$$
\eqalign{
\RR^{MN}(X) &\to \RR^{AB}(\Omega)\,{{\partial X^M}\over{\partial
                 \Omega^A}}\,{{\partial X^N}\over{\partial
                 \Omega^B}}     \cr
\RR(X)      &\to \RR(\Omega)      }
$$
in the equations of motion (15)  or, equivalently, Lagrangian
(14) . This gives
$$
I = - \k^2\int d^{4+d}X\sqrt{-G}\left[ R + F^{AB}(\Omega)\,
{{\partial X^M}\over{\partial \Omega^A}}\,{{\partial X^N}
\over{\partial \Omega^B}}\,G_{MN} - V(\Omega) \right]
                                                      \eqno(19) 
$$
where the target metric $F^{AB}(\Omega)$ and the potential
$V(\Omega)$ are defined as
$$
F^{AB}(\Omega)\equiv \RR^{AB}(\Omega)
$$
$$
V(\Omega)\equiv 2\,\RR(\Omega) \ .
$$
The identification of the new coordinates with the old ones,
$\Omega^A = X^A$, takes us back to the non-covariant theory.
The higher dimensional Lagrangian (19)  looks like a
$\s$-model coupled to gravity, but is certainly not of a
standard type. The derivatives of the scalar fields $\Omega^A$
appear non-polynomially in the action. We can bring it to a
polynomial form by introducing a set of auxiliary fields. In
particular, we need $4+d$ vector fields $b^A_{\ M}$ subject
to the equations of motion $b^A_{\ M}=\Omega^A_{\ ,\,M}\,$.
The easiest way to achieve this is to postulate the action
functional
$$
I = - \k^2\int d^{4+d}X\sqrt{-G}\left[ R + F^{AB}(\Omega)\,
h^M_{\ A}\,h_{MB} + \l^M_{\ A}\,(b^A_{\ M} -
\Omega^A_{\ ,\,M}) - V(\Omega) \right]                  \eqno(20) 
$$
with $h^M_{\ A}$ the inverse of $b^A_{\ M}$, and Lagrange
multipliers $\l^M_{\ A}$. That it is indeed equivalent
\eject\noindent
to (14)  is shown by inspecting the equations of motion. One
finds

$$
b^A_{\ M}=\Omega^A_{\ ,\,M}                          \eqno(21 a)
$$
$$
\l^M_{\ A}=2\,F^{BC}(\Omega)\,h^L_{\ B}h^M_{\ C}h_{LA}  \eqno(21 b)
$$
$$
R^{MN}= F^{AB}(\Omega)\,h^M_{\ A}h^N_{\ B}-{2\over
{d+2}}G^{MN}\left[ F^{AB}(\Omega)h^L_{\ A}h_{LB}-
{1\over 2}V(\Omega)\right]                          \eqno(21 c)
$$
$$
{{\partial V}\over{\partial \Omega^A}}={{\partial
F^{BC}}\over{\partial \Omega^A}}h^M_{\ B}h_{MC}-{1\over 2}
G_{MN}\,G^{MN}_{\ \ \ \ ,L}\,\l^L_{\ A}+\l^M_{\ A,M}\ .   \eqno(21 d)
$$
\vskip.6cm
\noindent
The auxiliary fields $b^A_{\ M}$ and $\l^M_{\ A}$ are fully
expressed in terms of $G_{MN}$ and $\Omega^A$, and carry no
degrees of freedom. The equation (21 d), obtained by varying
the action (20)  with respect to $\Omega^A$, is not an
independent equation of motion. It is easily shown to follow
from the Bianchi identities $(R^M_{\ N}-{1\over 2}\d^M_{\ N}
R)_{;M}\equiv 0$ and (21 a--c). It turns out then that the
content of the theory is fully contained in (21 c) with
$h^M_{\ A}=\partial X^M / \partial \Omega^A\,$. After the
spacetime coordinates are chosen to fix $\Omega^A=X^A$, the
equations of motion boil down to (15) , as expected.

The theory given by (20)  or, equivalently, (19)  differs in
some aspects from the geometric $\s$-models of reference
[2], and, in that respect, from the model of references
[3] and [4]. First, the equations of motion (21 ) do not
admit the topologically trivial solution $\Omega^A=0$
representing a non-geometric sector of the theory. Second,
although our target metric $F^{AB}(\Omega)$ has vanishing
$F^{\a B}\,$ ($\a=0,...,3$) components, we still need all
$4+d$ fields $\Omega^A$. In ordinary geometric $\s$-models,
the rank of the Ricci tensor $\RR_{MN}$ determines the
number of necessary scalar fields. This is why we needed
only $d$ out of $4+d$ scalar fields in the model defined
by (1). Here, we have to retain all the components $\Omega^A$,
in particular $\Omega^0$ whose vacuum value $\Omega^0=X^0$ is
time dependent. Still, this is a pure coordinate time
dependence which can easily be gauged away.

\section{6. Concluding remarks}

We have applied the ideas of geometric $\s$-models [2] to
solve the cosmological constant problem of Kaluza-Klein
theories. This kind of approach is not new in literature.
The authors of references [3] and [4] have demonstrated
how scalar fields in the form of a $\s$-model can
trigger spontaneous compactification. It turned out, however,
that their model failed to give massless gauge fields after
the dimensional reduction. We have rewritten this theory
in terms of a geometric $\s$-model, thereby bringing it
to a suggestive form. It was not difficult then to realize
which kind of modification would reconcile the masslessness
of the gauge fields with the zero value of the cosmological
constant. In section 3, the modified theory has been proven
to contain no massive gauge fields. The effective
4-dimensional theory has been obtained by averaging the linearized
$(4+d)$-dimensional equations of motion over the internal
coordinates.

In search for the simple Lagrangian of the theory, we had to
abandon the model of section 3, and look for another
modification. We have found an action functional whose
equations of motion differ from those of section 3 by the presence
of a term proportional to $(G_{MN}-\GG_{MN})$, but
retain all their good features. The linearized effective
4-dimensional equations of motion turned out to contain
the standard Einstein, Yang--Mills and Klein--Gordon sectors.
In addition, the scalar excitations of the internal manifold, in
particular their zero mode, have been shown to have masses of the order
of the Planck mass.

In section 5, we have covariantized our theory. A set of $4+d$ scalar
fields has been introduced in a purely geometric manner. The generally
covariant theory turned out to be of the form of a non-standard
$\s$-model with non-polynomial dependence on the scalar field derivatives.
We have demonstrated how the introduction of auxiliary fields brings it
to a polynomial form. Compared to geometric $\s$-models of reference [2],
and, in that respect, to the model of [3] and [4], our theory exhibits
some differences. In particular, the number of scalar fields needed for
the covariantization does not match the rank of the vacuum value of the
Ricci tensor.

In the course of our analysis, it became obvious that the form of the
dynamics was chosen from a variety of possibilities. To decide upon one,
we have to study its physical implications. The first thing one should
check is the general stability of the vacuum state. If the dynamics of
the theory does not support the stable $M^4\times B^d$
vacuum configuration, it should be abandoned. If it does, we still have to
compare the implications of the interacting theory with the known
results. In search for a realistic theory of the kind, we could also further
develop the idea of [2] to give fermions a pure geometric origin. In
particular, it would be more in the spirit of geometric $\s$-models if we
chose our vacuum metric in the form of a localized, particle-like field
configuration which only asymptotically approaches $M^4\times B^d$. The
corresponding theory of the type considered in this paper might turn out
to be more promising.

\section{Acknowledgments}

This work has been supported in part by the Serbian Research Foundation,
Yugoslavia.

\vfill\eject

\section{References}

\item{[1]} Kaluza T 1921 {\it Sitzungsber. Preuss. Akad. Wiss. (Berlin)
           Math. Phys.}\hfill\break {\bf K1} 966  \hfill\break
           Klein O 1926 {\it Z. Phys.} {\bf 37} 895  \hfill\break
           Appelquist T, Chodos A and Freund P G O 1987 {\it Modern
           Kaluza--Klein \hfill\break theories (Addison--Wesley)}

\item{[2]} Vasili\'c M 1998 {\it Class. Quantum Grav.} {\bf 15} 29
\item{[3]} Omero C and Percacci R 1980 {\it Nucl. Phys.}
           B {\bf 165} 351
\item{[4]} Gell-Mann M and Zwiebach B 1984 {\it Phys. Lett.}
           {\bf 141B} 333
\item{[5]} Salam A and Strathdee J 1982 {\it Ann. Phys. NY} {\bf 141} 316
\item{[6]} Vasili\'c M 1994 {\it Nuovo Cimento} {\bf B109} 1083

\vfill\eject\end